%% file: 0.main.tex
\documentclass[lettersize,journal]{IEEEtran}
\usepackage[utf8]{inputenc} 
\usepackage[T1]{fontenc}
\usepackage{url}
\usepackage{ifthen,hyperref}
\usepackage[cmex10]{amsmath}
\usepackage{array} 

\usepackage{enumerate}
\usepackage{amssymb}
\usepackage{amsmath} 
\usepackage{xcolor}

\usepackage{algorithm}
\usepackage{algorithmicx}
\usepackage{algpseudocode} 

\usepackage{hhline}
\usepackage{mathrsfs}
\usepackage{bm}
\usepackage{tikz}
\usetikzlibrary{arrows}
\usepackage[caption=false]{subfig}
\usepackage{graphicx, booktabs, multirow}
\usepackage{epstopdf}
\usepackage{multirow}
\usepackage{tabularx} 
\usepackage{booktabs}
\usepackage{cite}

\usepackage{pifont}

\definecolor{colorhkust}{RGB}{20,43,140}
\definecolor{colortsinghua}{RGB}{116,52,129}
\definecolor{color1}{RGB}{128,0,0}

\usepackage{amsthm}

\usepackage{flushend}

\theoremstyle{definition}

\theoremstyle{remark}

\begin{document}
      \title{AI Flow at the Network Edge}
      \author{Jiawei Shao and Xuelong Li,~\IEEEmembership{Fellow,~IEEE}
      \thanks{
     The authors are with the Institute of Artificial Intelligence (TeleAI), China Telecom, China (E-mail: shaojw2@chinatelecom.cn, xuelong\_li@ieee.org). The corresponding author is Xuelong Li.
     }
     }
        \maketitle
\begin{abstract} 
Recent advancements in large language models (LLMs) and their multimodal variants have led to remarkable progress across various domains, demonstrating impressive capabilities and unprecedented potential.
In the era of ubiquitous connectivity, leveraging communication networks to distribute intelligence is a transformative concept, envisioning AI-powered services accessible at the network edge. 
However, pushing large models from the cloud to resource-constrained environments faces critical challenges.
Model inference on low-end devices leads to excessive latency and performance bottlenecks, while raw data transmission over limited bandwidth networks causes high communication overhead.
This article presents \emph{AI Flow}, a framework that streamlines the inference process by jointly leveraging the heterogeneous resources available across devices, edge nodes, and cloud servers, making intelligence flow across networks.
To facilitate cooperation among multiple computational nodes, the proposed framework explores a paradigm shift in the design of communication network systems from transmitting information flow to intelligence flow, where the goal of communications is task-oriented and folded into the inference process.
Experimental results demonstrate the effectiveness of the proposed framework through an image captioning use case, showcasing the ability to reduce response latency while maintaining high-quality captions.
This article serves as a position paper for identifying the motivation, challenges, and principles of AI Flow.

\end{abstract}

\input{1.Introduction}

\input{2.System_overview}

\input{3.Cooperative_inference}

\input{4.Inference_speedup}

\input{5.Experiments}

\input{6.Conclusions}

\bibliographystyle{ieeetr}
\bibliography{ref}

\noindent{\bf{Jiawei Shao}} [S'20-M'24] (shaojw2@chinatelecom.cn) is a Research Scientist at the Institute of Artificial Intelligence (TeleAI), China Telecom. 
He received his Ph.D. at the Hong Kong University of Science and Technology.
\\
{\bf{Xuelong Li}} [M'02-SM'07-F'12] (xuelong\_li@ieee.org) is the CTO and Chief Scientist of China Telecom, where he founded the Institute of Artificial Intelligence (TeleAI) of China Telecom.

\end{document}

%% file: 1.Introduction.tex
\section{Introduction}

\begin{figure*}[t!]
    \centerline{\includegraphics[width=0.99\linewidth]{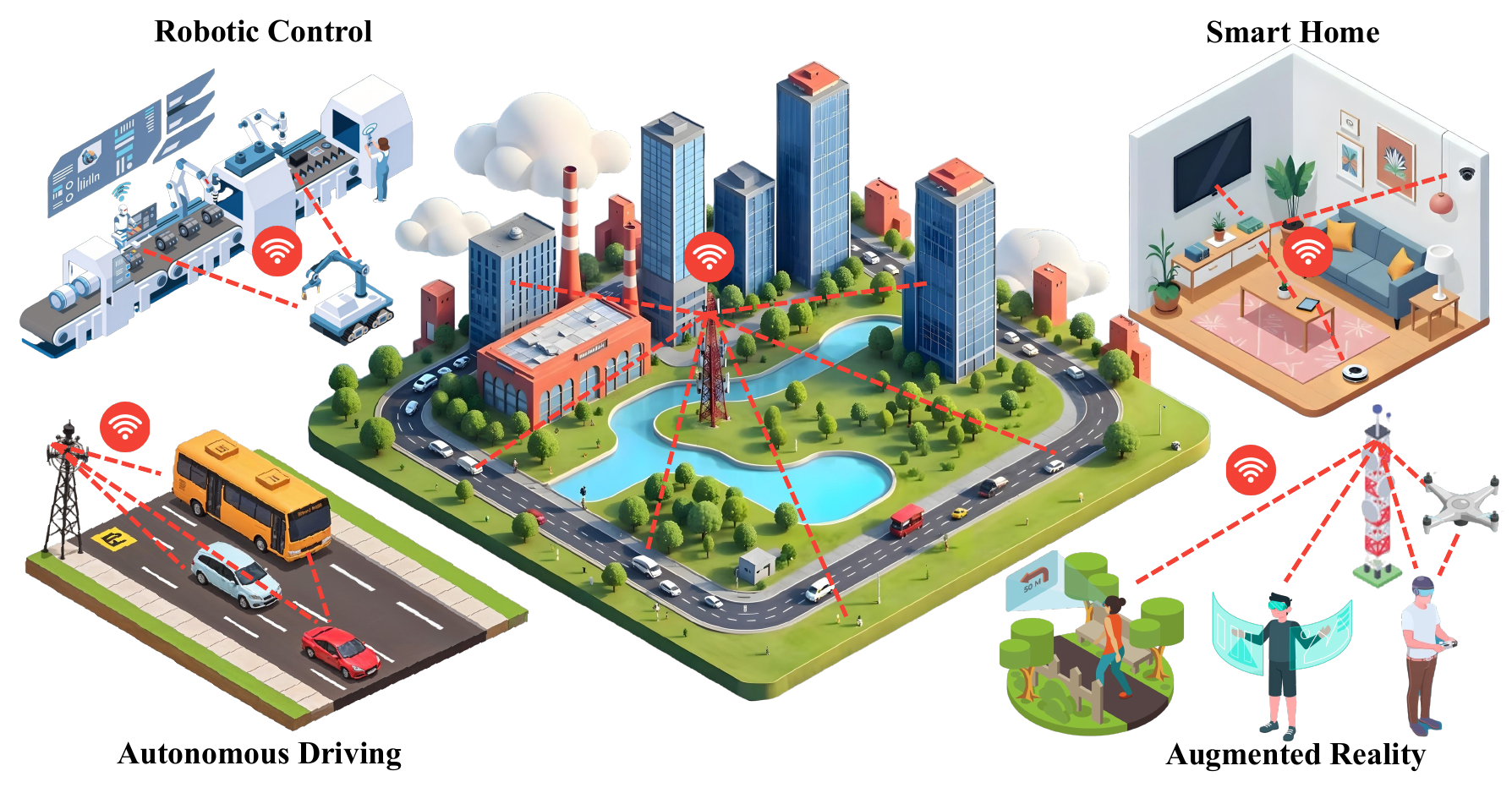}}
    \caption{Typical intelligence applications at the network edge.
    }
    \label{fig:applications}
\end{figure*}

We are witnessing a transformative era in the field of artificial intelligence (AI) with groundbreaking technologies \cite{brown2020language,liu2024visual}. 
LLMs, like ChatGPT and PaLM, have demonstrated remarkable capabilities in natural language understanding, generation, and reasoning. 
Going beyond language, multimodal foundation models, like CLIP and Llama,  have showcased exceptional performance in cross-modal perception, understanding, and interaction tasks.
These advancements led to their adaptations in a broad spectrum of application domains, including embodied robotics, augmented reality, and autonomous driving \cite{duan2022survey,king2024sasha,zhang2024empowering,xu2024drivegpt4}.
Most recently, edge devices (mobile phones, smart wearables, and IoT sensors) are becoming increasingly widespread, and sensory data are easy to access.
This has sparked a surge of interest in pushing intelligence applications from the central cloud to the network edge \cite{shen2024large,letaief2022Edge}.
As visualized in Fig. \ref{fig:applications}, the future envisions a scenario where AI technologies are seamlessly integrated into various aspects of daily life.

While offering exciting opportunities, deploying large models at the network edge faces new challenges.
In contrast to cloud intelligence, performing AI tasks at the network edge differs significantly in many cases.
First, the fundamental difference is the available resources. 
Edge devices are usually equipped with limited computation resources, whereas cloud servers possess a large number of powerful processing units. 
As a result, on-device model inference struggles to support real-time responses since the large foundation models demand intensive computation \cite{shao2020communication}.
Besides, edge devices are often wirelessly connected.
The limited uplink bandwidth and the highly dynamic nature of wireless channels hinder transmitting large volumes of data collected by edge devices to cloud servers in real time.

Existing studies on edge AI typically focus on small models while overlooking the complexities and resource demands of large-scale transformer-based models.
To address this gap and efficiently provide diverse intelligent services, we propose \emph{AI Flow}, a framework that optimizes the deployment and execution of both small and large models.
Specifically, AI Flow streamlines the inference process by jointly leveraging the heterogeneous computational resources available across devices, edge nodes, and cloud servers.
This framework distributes the inference workload across different network layers and adapts to dynamic network conditions.
To facilitate cooperation among multiple computational nodes, AI Flow provides a paradigm shift in the design of communication network systems from transmitting information flow to intelligence flow.
The goal of communications needs to be folded into the inference process.
Instead of sending raw data from edge devices to servers for processing, the intelligence flow features a task-oriented property, where edge devices extract only critical features from the raw sensory data and discard redundant information to reduce communication overhead. 
To summarize, the advantages of the AI Flow framework are manifold.
\begin{itemize}
\item Ubiquity: The widespread deployment of AI capabilities at the edge makes intelligent responses possible wherever there is network access. 
This ensures that various devices benefit from AI-driven insights without the need for constant connectivity to central servers.
\item Adaptivity: The framework distributes inference tasks according to available computational resources, dynamic network conditions, and task requirements. 
Adaptively scheduling task execution maintains performance in fluctuating environments.
\item Low-latency: By prioritizing data processing close to the source, the delay in transmitting information to powerful servers is minimized. 
This provides a quicker end-to-end response and achieves a smoother user experience.
\end{itemize}
The rest of the article is organized as follows.
Section \ref{sec:applications} introduces typical applications at the network edge, and \ref{sec:2.system} provides a system overview of AI Flow.
Two types of enabling techniques, namely, cooperative inference and model inference speedup, have been elaborated in Section \ref{sec:3.cooperative} and Section \ref{sec:4.inference}, respectively.
Section \ref{sec:5.exp} presents a case study to evaluate the effectiveness of the proposed AI Flow framework.
Finally, we conclude this article and point out future research opportunities in Section \ref{sec:6.conclusion}.

\section{Typical Intelligence Applications at the Network Edge}
\label{sec:applications}

Large models can be directly applied or fine-tuned to a broad range of tasks. 
As illustrated in Fig. \ref{fig:applications}, this part will focus on four promising use cases: robotic control, smart home, augmented reality, and autonomous driving.

Robotic control \cite{duan2022survey} is crucial for optimizing the interaction between robots and the environment.
Vision-language models (VLMs) enhance the perception capabilities of robots.
For instance, by combining three-dimensional point clouds with visual-language features from a VLM, a three-dimensional map of the physical world can be created. 
Integrating audio signals and haptic features with visual data enables embodied robotics to navigate using multimodal objectives.
With remarkable generalization capabilities, LLMs allow robots to comprehend human instruction or complicated environments, enabling them to perform embodied reasoning and break down complex tasks into actionable steps.
Nevertheless, centralized robotic control involves massive streaming video upload, which could overwhelm the wireless networks.

Smart homes \cite{king2024sasha} benefit from the integration of multimodal large language models (MLLMs), which utilize diverse sensory data from the home environment. 
For instance, MLLMs can intelligently adjust air conditioner settings by integrating temperature data and voice commands.
Furthermore, the channel state information (CSI) obtained from WiFi sensing provides extra information for localization.
This allows MLLMs to make context-aware decisions, such as turning lights on or off based on the predicted trajectory of humans inferred from the CSI.
However, the large models demand intensive computational resources. 
While heterogeneous devices in smart homes provide distributed computing power, they remain insufficient for real-time inference due to the limitations of network bandwidth and the difficulty in synchronization.

Augmented reality \cite{zhang2024empowering} enhances user interactions with their surroundings through glasses, headsets, and other smart wearables.
These devices overlay digital information, guided by generative models, directly into the field of vision. 
For visually impaired individuals, AI glasses can highlight important visual features such as object edges, provide navigational cues, or display relevant contextual information about the environment, greatly enhancing spatial awareness. Additionally, LLM-powered devices can streamline human communication by translating spoken language into clear, concise text displayed in real-time. 
However, the need to sense the environment and process large volumes of multimodal data makes it impractical for resource-constrained headsets.
This necessitates the development of more efficient processing techniques and the optimization of data transmission at the network edge.

Autonomous driving \cite{xu2024drivegpt4} has great promise to revolutionize transportation.
VLMs enable a combination of scene description and hierarchical planning.
Typical methods leverage video from a front-view camera to make driving decisions regarding speed and angle.
With LiDAR sensor data, more accurate perception can be achieved than with visual data alone, as camera imagery can be impacted by weather conditions. 
Moreover, LLMs can use textual data, such as traffic and weather reports, to optimize route planning.
Additionally, LLMs interacting with passengers help them to understand the motion control process by explaining each driving decision. 
However, autonomous driving has stringent latency requirements. High latency between vehicles and the data center poses a risk to safety.
Addressing this issue is crucial for the successful implementation of autonomous driving systems.

%% file: 2.System_overview.tex
\begin{figure*}[ht]
    \centerline{\includegraphics[width=0.69\linewidth]{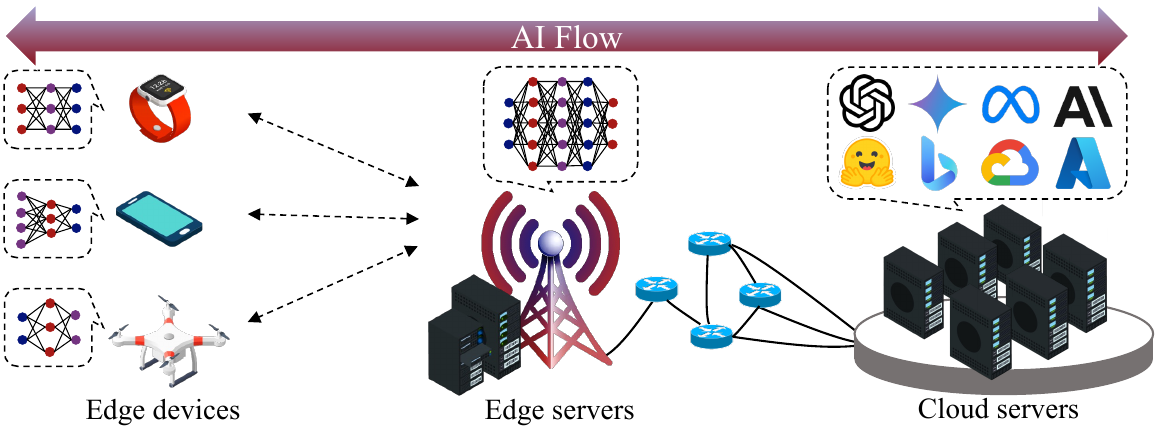}}
    \caption{A system overview of the AI Flow framework.
    }
    \label{framework}
\end{figure*}
\section{System Overview}
\label{sec:2.system}

This section outlines the modules, inference paradigms, and key design considerations in AI Flow.

\subsection{Modules}
The architecture, as illustrated in Fig. \ref{framework}, comprises the following three modules.
\subsubsection{Edge devices}
The edge devices denote the end-user equipment, such as smartphones, wearables, and IoT sensors, that collect data and interact with users directly.
They are essential for initial data capture, preliminary processing, and information transmission. 
As these devices are often the first touchpoint in data processing, they play a critical role in determining the responsiveness and effectiveness of the AI Flow framework.
However, constrained by limited computing and energy resources, these devices cannot perform intensive computation or storage-heavy tasks.
This limitation necessitates a cooperative processing arrangement with edge or cloud servers.

\subsubsection{Edge servers} 
The edge servers provide processing, storage, and other resources at an edge location.
Typical edge servers are deployed at base stations and cell towers that are directly connected to edge users and devices.
By processing data closer to the source, edge servers can significantly reduce the round-trip time without sending data to the remote cloud server. This enables a faster response time for latency-sensitive applications, such as real-time video streaming, gaming, and autonomous vehicles.
Besides, edge computing allows for the distribution of processing and storage resources across multiple edge devices and servers, fully leveraging the heterogeneous resources at the network edge.

\subsubsection{Cloud servers} 
The cloud servers provide powerful computing resources and vast storage capacities essential for handling complex and resource-intensive tasks.
They also have access to a broad range of AI services and tools that can be allocated based on demand. 
In addition, the cloud center plays a critical role in network management, capable of scheduling data flow and integrating information from various edge nodes for more sophisticated analytics.
However, due to the physical distance between the cloud and end devices, relying solely on cloud servers may introduce higher latency and consume more bandwidth compared to edge computing.

\subsection{Inference Paradigms}
There are three types of cooperative schemes in AI Flow, including, on-device inference, device-edge cooperative inference, and device-edge-cloud cooperative inference.

\subsubsection{On-device inference}
In this scheme, the devices complete some simple inference tasks locally and independently based on the built-in AI chip. 
Due to limitations in storage and computational resources, model light-weighting techniques are widely adopted to reduce storage and computational costs.
In addition, hardware accelerators, such as neural processing units and field programmable gate arrays, are utilized to enhance processing capabilities and energy efficiency.
These enable the scalable deployment of AI models, making on-device inference a widely adopted solution.

\subsubsection{Device-edge cooperative inference}
While on-device inference is suitable for basic tasks, it struggles to handle complex tasks that require larger AI models, leading to excessive latency. 
Edge computing offers supplementary computational resources as a promising solution.
This setup allows devices to either send raw data directly to edge servers for inference or to perform preliminary computations locally and then send the intermediate results to the edge servers for further processing.
Instead of uploading data to a central server far away, the proximity to data sources reduces latency, saves bandwidth, and improves system responsiveness.

\subsubsection{Device-edge-cloud cooperative inference}

For tasks that require extensive computational resources, massive data storage, or external knowledge, the integration of the device-edge setup and the cloud infrastructure creates a hierarchical architecture.
Initial data processing begins at the network edge.
Then the cloud provides further computational power and access to knowledge bases to refine the results.
For example, retrieval-augmented generation requires abundant extra contextual data to alleviate the hallucination issue.
Re-identification tasks need large-scale databases to perform intensive search operations.
With sufficient computational power, the beam search maintains multiple hypotheses to find the most probable sequence of tokens.
Additionally, the cloud can also aggregate information from multiple edge nodes to gain a more comprehensive view. 
This cooperative inference scheme leverages the respective advantages of devices, edge, and cloud to enable AI services that are beyond the capabilities of any single tier.

\subsection{Key Design Considerations}

In designing the AI Flow framework, several critical considerations must be addressed to ensure performance, efficiency, and user experience.
Two questions that need to be answered are as follows:

\textit{1) How to enable efficient cooperation?} 
Given the constrained computing capabilities of edge devices, extra computational support is essential to complete inference tasks. 
A significant challenge in this process is the substantial communication latency incurred during the transmission of sensory data or intermediate results. 
Additionally, variations in wireless connectivity between edge devices and edge servers, along with changes in routing from the edge to the cloud, can further delay response times.

\textit{2) How to speed up model inference?}
The most notable characteristic of large models is the massive number of parameters, bringing significant computational demands.
The computational complexity of transformer-decoder models increases quadratically with the token length.
While methods like key-value (KV) cache mitigate this challenge by avoiding recomputation, substantial memory consumption limits the scalability.
This results in frequent memory overflows and creates new challenges for memory management.
In addition, real-time inference in large models is often slow due to their autoregressive nature.


To address the above questions, two types of enabling techniques are introduced in the following Section \ref{sec:3.cooperative} and Section \ref{sec:4.inference}.

%% file: 3.Cooperative_inference.tex
\section{Cooperative Inference}
\label{sec:3.cooperative}

Providing intelligence services at the edge devices typically requires transmitting raw data to the remote servers for processing.
When the raw data collected by the edge sensors are large in size, this framework leads to high bandwidth usage and delay.
Cooperative inference is a promising solution that offloads the inference task to multiple computational nodes.
Edge devices can first perform data pre-processing to discard redundancy and determine what information is important for transmission.
A well-designed cooperative strategy can strike a balance between the communication overhead and the on-device computation to reduce the end-to-end system latency.

\begin{figure}[t!]
    \centerline{\includegraphics[width=0.965\linewidth]{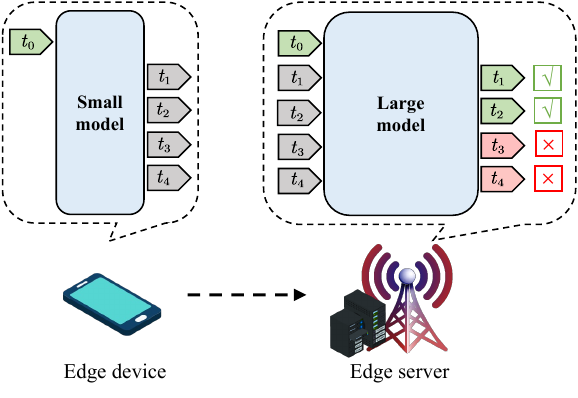}}
    \caption{Cooperation between small and large models for edge inference based on speculative decoding. A small model and a large model are deployed at an edge device and an edge server, respectively. 
    In this example, the small model generates four draft tokens ($t_{1},t_{2},t_{3},t_{4}$) and sends them to the large model for verification. 
    The first two tokens pass the verification while the last two fail.
    }
    \label{fig:speculative}
\end{figure}

\begin{figure*}[t!]
    \centerline{\includegraphics[width=0.9\linewidth]{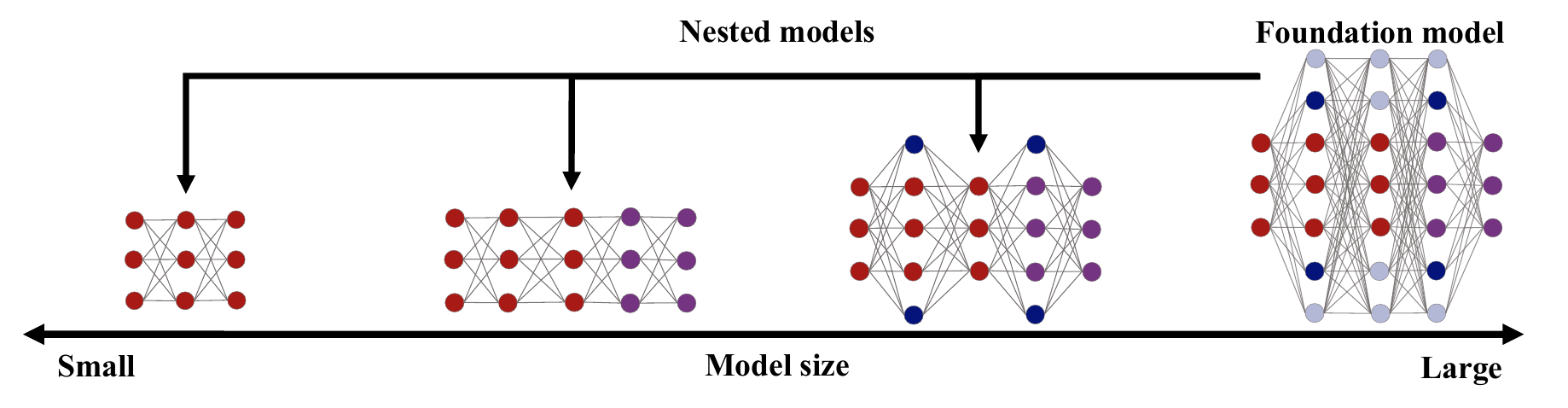}}
    \caption{An illustration of the nested neural network. A large foundation model contains sub-models of different sizes.
    These sub-models share parameters by being nested within the larger ones.
    }
    \label{fig:nested}
\end{figure*}

\subsection{Theoretical Insights}

The critical problem in cooperative inference is identifying task-relevant information from raw data and discarding redundancy to reduce communication overhead. 
Based on the information bottleneck principle, cooperative inference can be characterized by its ability to filter and compress input data.
Only essential features should be sent to centralized systems for further analysis, where the amount of information is upper bounded by the mutual information $I(X;Y)$ between the input data $X$ and the target output $Y$. 
This characteristic implies that significant data compression is achievable without degrading the inference performance.
Ideally, denoting the random variable $Z$ as the minimal information that maintains the inference performance, such $Z$ satisfies that the mutual information terms $I(X;Z)$, $I(Y;Z)$, and $I(X;Y)$ are equal \cite{alemi2022deep,shao2021learning}. 
Assuming the mapping from $X$ to $Z$ is deterministic, the minimal communication overhead is the entropy of $Z$.
To reduce the end-to-end latency in cooperative inference, the key lies in adopting a small model for compact feature extraction at the edge device.

\subsection{Enabling Techniques}

Many techniques are available to support such cooperative inference architecture, where a small part of the computation task is completed at the edge side, which can be seen as a feature extraction process, and more computation-intensive operations are supported by the cloud.

Split inference is such a method, which splits a model into two parts and deploys one part at the device and the other part at the server.
Instead of sending the raw data, split inference allows sending the intermediate activations at the split point to the server.
A variant of split inference is the early exiting method. 
By selecting multiple exit points in the backbone of the model and augmenting additional side branch networks, this architecture allows input samples to exit at different points.
The powerful server leverages the corresponding branch network for further processing.
For transformer-based models, a similar idea is to deploy the near-input embedding layer at the device and offload the computational-intensive transformer blocks at the server.
Such deployment extracts initial contextual embeddings that are generalizable to multiple downstream tasks.

The synergy of large and small models can also support cooperative inference.
As illustrated in Fig. \ref{fig:speculative}, one such approach is speculative decoding \cite{leviathan2023fast}.
Typical LLMs and VLMs need to generate tokens in an autoregressive manner that outputs tokens iteratively.
This process is slow and computationally intensive.
Speculative decoding facilitates the simultaneous decoding of multiple tokens.
It introduces a novel inference paradigm where a small model deployed at the edge device drafts multiple tokens once. 
Then a large model at the server verifies and corrects these tokens in parallel. 
When the acceptance rate of draft tokens by the large model is high, significant latency reduction can be achieved compared to the traditional server-only inference.
This is due to the low cost of generating draft tokens by a small model, coupled with the low overhead of transmitting these tokens.

%% file: 4.Inference_speedup.tex
\section{Model Inference Speedup}
\label{sec:4.inference}

Powerful AI models have achieved great success in various fields. 
Meanwhile, they are experiencing rapid growth in model size.
The substantial computational load of large foundation models presents a significant bottleneck, especially for model inference on resource-constrained devices.
It is widely recognized that many popular deep neural network architectures are over-parameterized.
This indicates that maintaining high performance does not necessarily require large models.
Therefore, there is an opportunity for model compression and inference speedup without sacrificing effectiveness.

\subsection{Theoretical Insights}

The typical model training involves finding weights such that the maximum information pertaining to the target output propagates from an input to the network output.
Model compression can be characterized as minimizing the information redundancy between adjacent layers while maintaining the relevance between the activations and the target output \cite{dai2018compressing}.
The information bottleneck principle provides a convenient mechanism for penalizing information redundancy in data processing.
Denote the hidden layer activations at layer $i-1$ and layer $i$ as $Z_{i-1}$ and $Z_{i}$, respectively.
For each layer $i$, the information bottleneck objective targets minimizing the mutual information between $Z_{i-1}$ and $Z_{i}$, while maximizing the mutual information between $Z_{i}$ and the target output $Y$.
From the perspective of information capacity (IC), the efficiency of a model is characterized by the amount of preserved information divided by the number of parameters.
To reduce the model size and speed up inference, the key lies in identifying and removing redundant information propagation across layers.


\subsection{Enabling Techniques}

There are many empirical successes in speeding up model inference through compression techniques.
Model parameter pruning is devoted to removing unimportant components.
By aggregating useful information into a subset of activations, other neurons or channels containing task-irrelevant information can be pruned without performance loss.
Similarly, low-rank factorization decomposes parameter matrices into lower-dimensional components, eliminating redundant information and reducing the model size.
Neural architecture search leverages reinforcement learning to design model architectures that balance performance with computational efficiency.
Besides, model quantization decreases the informativeness of activations by transforming high-precision weights into lower-bit float or integer values.
Knowledge distillation trains a small student model to mimic the functionality of a large teacher model.
Alternatively, the adoption of explainable AI \cite{liatsas2024xai} forecasts the workload and identifies critical components in the neural architecture.
This enables model designers to adjust the tradeoff between model complexity and performance drop.

Besides permanently removing some parts of the model, dynamic neural networks are capable of adjusting their structure or parameters on the fly, depending on the input they receive.
In particular, Mixture-of-Experts (MoE) characterizes different parts of a model as experts specializing in different tasks. Only pertinent experts are engaged for a given input, striking a balance between computational efficiency and model capability.
Similarly, Mixture-of-Depths (MoD) chooses to either apply computation to a token or pass it through a residual connection.
Such routing emphasizes how individual tokens pass through different numbers of layers, or blocks, through the depth of the transformer.
Furthermore, nested neural networks provide scalable and adaptive performance capabilities.
As shown in Fig. \ref{fig:nested}, a large foundation model consists of a series of sub-models, which share parameters through a nesting structure.
These sub-models are arranged hierarchically, such that smaller sub-models form the core of larger ones. 
Each sub-model varies in size and can solve the same task as the foundation model. 
This hierarchical design enables efficient storage and flexible model deployment to meet diverse budget constraints.

Furthermore, there are complementary techniques that can enhance the inference efficiency of large models.
During the autoregressive generation, KV pairs from previous tokens are stored in a cache to assist in generating future tokens. 
However, given a long input sequence, the cache size increases rapidly, leading to long inference latency.
Popular optimization techniques include quantizing the KVs and evicting the less informative KVs. 
These approaches allow for faster token generation by strategically compressing the cache.
Additionally, some novel low-complexity attention mechanisms, such as the Informer, can be adopted to accelerate inference \cite{gort2024forecasting}.

%% file: 5.Experiments.tex
\begin{figure}[t!]
    \centerline{\includegraphics[width=0.95\linewidth]{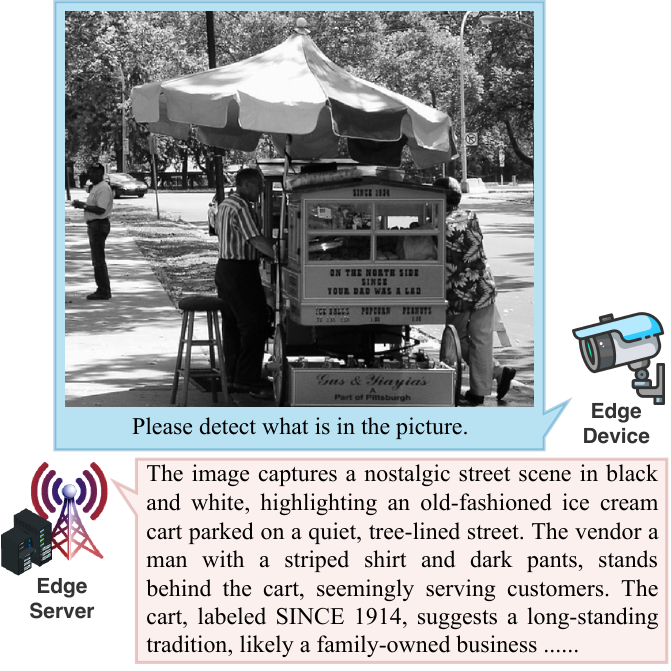}}
    \caption{An illustration of the image captioning task.
    An edge device captures real-time images of certain places and then asks for Visual-LLM service to detect objects and identify ongoing events.
    }
    \label{fig:interacte}
\end{figure}

\section{Case Study: Speculative Decoding for Edge Inference}
\label{sec:5.exp}

This section provides performance evaluations to demonstrate the performance of AI Flow, which leverages the speculative decoding approach to enhance the efficiency of cooperative inference.

\subsection{Setup}
Consider a device-edge cooperative inference system.
The edge device is equipped with a GeForce RTX 4090, and the edge node is a powerful GPU cluster.
They are connected by wireless channels with data rates ranging from 500 KB/s to 2 MB/s.
We evaluate the performance of speculative decoding on an image captioning task served at the edge, as shown in Fig. \ref{fig:interacte}.
The experiments are conducted on the Vehicles-OpenImage dataset, which is a subset of the OpenImage dataset containing 627 images of various vehicle classes in open space. 
The small and large models are selected from the InternVL2 series, which is a high-scalable VLM family that includes models ranging from a 1B model to a significantly more powerful 108B model. 
In the experiment, a small InternVL2-2B model is deployed at the device to generate draft tokens. Meanwhile, a large InternVL2-26B model is deployed on the edge server. 
For the baseline method, known as server-only inference, the edge device transmits raw images directly to the edge server. The server then uses the large model for inference.
We focus on the response latency in the experiment and select the time per output token (TPOT) as the metric.
To maintain a fair comparison, the large model in speculative decoding corrects all inconsistent tokens generated by the small model.
This guarantees that the final output of speculative decoding matches exactly with the output of server-only inference.

\begin{figure}[t!]
    \centerline{\includegraphics[width=0.99\linewidth]{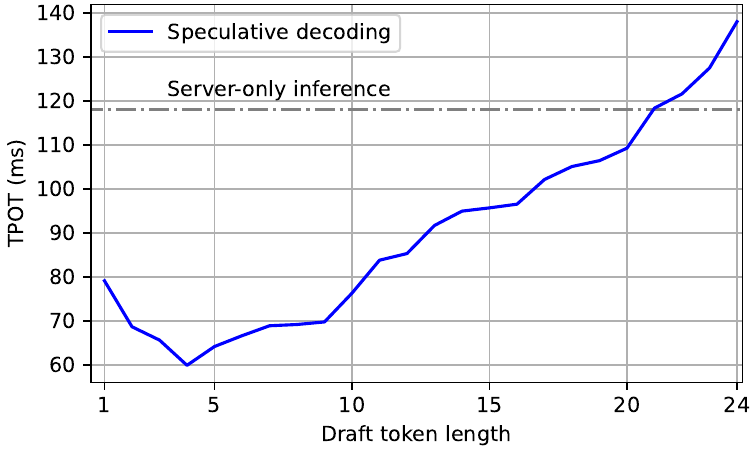}}
    \caption{Time per output token (TPOT) as a function of the draft token length.
    }
    \label{fig:spec_results}
\end{figure}

\subsection{Results}
As shown in Fig. \ref{fig:spec_results}, speculative decoding significantly reduces the TPOT compared with the baseline method.
In particular, the optimal draft token length in this experiment is 4, where the inference speed is approximately double that of the baseline.
This optimal length balances the acceptance rate of draft tokens with their quantity, maximizing the number of tokens that pass verification at the edge server.
When the draft token length is set below 4, there is a slight increase in response latency.
This occurs because generating fewer draft tokens does not fully leverage the predictive capabilities of the smaller model on the edge device. 
In contrast, setting the draft token length above 4 leads to an increase in response latency. 
This is because the smaller model on the edge device produces more errors with longer token sequences, resulting in a lower rate of token acceptance.
Specifically, once the token length exceeds 22, the overall latency is higher than that of the baseline.
Therefore, carefully selecting the optimal draft token length has the potential to enhance the performance gains of speculative decoding in cooperative inference systems.

\subsection{Discussion}

In this case study, we focus on the effectiveness and efficiency of the speculative decoding method for cooperative inference. 
Empirical results show that the device-edge cooperative scheme can largely reduce response latency compared to server-only inference.
It is important to note that, as discussed in Section V, many other inference speedup techniques, such as model compression, dynamic neural networks, and KV cache optimization, are available. These techniques are complementary to the cooperative inference scheme and can be combined with speculative decoding to achieve even greater efficiency. 
For instance, model compression could further reduce the computational load on both the device and the server, while KV cache optimization could significantly improve the speed of token verification.
Exploring the interplay between these techniques represents a promising direction for future research.
By combining multiple acceleration strategies, it is possible to achieve significant performance gains in edge inference scenarios.

%% file: 6.Conclusions.tex
\section{Conclusions and Future Work}
\label{sec:6.conclusion}

This article introduced AI Flow, a novel framework designed to optimize the inference process by effectively utilizing heterogeneous resources across devices, edge nodes, and cloud servers. 
To facilitate cooperation among multiple computational nodes, we advocated a paradigm shift in the design of communication network systems from reliably reconstructing the raw data to transmitting the most task-relevant information.
This allows edge devices to extract only critical features from the raw sensory data and discard redundant information to reduce communication overhead. 
A proof-of-concept case study demonstrated the effectiveness of the proposed framework through an image captioning use case, showcasing the ability to reduce response latency while maintaining high-quality captions.
Several areas could be explored for future advancements.

\textbf{Security and privacy considerations:} Our framework prioritizes inference efficiency, but security and privacy are also important in edge systems.
Integrating techniques such as homomorphic encryption and differential privacy mechanisms can reduce attack surfaces and protect sensitive information against unauthorized access or data breaches.

\textbf{Software-hardware co-design:} 
While we center on software algorithms, bringing hardware and software closer could lead to further optimizations.
High-throughput generation engines can be flexibly configured by aggregating memory and computation from heterogeneous resources.
Besides, creating specialized processors tailored to AI operators like matrix multiplication and convolutions also speeds up model execution.

\textbf{Scalability and stability:} 
Concurrent requests from a large number of edge devices may strain the system. 
Strategies for load balancing, workload scheduling, and efficient resource allocation need further investigation.
Redundancy and failover mechanisms should also be incorporated to ensure system resilience and maintain high availability.